\documentclass[3p,times,twocolumn]{elsarticle}
\usepackage{ecrc}
\volume{00}
\firstpage{1}
\journalname{Nuclear and Particle Physics Proceedings}
\runauth{G.G.~Barnaf\"oldi et al.}
\jid{nppp}
\jnltitlelogo{Nuclear and Particle Physics Proceedings}
\usepackage{amssymb}
\usepackage[figuresright]{rotating}
\usepackage{xcolor}

\begin{document}

\begin{frontmatter}



\dochead{}

\title{First Results with HIJING++ in High-Energy Heavy-Ion Collisions}


\author[1]{Gergely G\'abor Barnaf\"oldi}
\ead{barnafoldi.gergely@wigner.mta.hu}
\author[1,2]{G\'abor B\'\i r\'o}
\author[1,3,4,5]{Miklos Gyulassy}
\author[1,2]{Szilveszter Mikl\'os Harangoz\'o}
\author[1]{P\'eter L\'evai}
\author[5]{Guoyang Ma} 
\author[2]{G\'abor Papp}
\author[4,5]{Xin-Nian Wang}
\author[5]{Ben-Wei Zhang}
\address[1]{Wigner Research Centre for Physics of the Hungarian Academy of Sciences, 29-33 Konkoly-Thege Mikl\'os Str, H-1121 Budapest, Hungary}
\address[2]{E\"otv\"os Lor\'and University, 1/A P\'azm\'any P. S\'et\'any, H-1117, Budapest, Hungary} 
\address[3]{Pupin Lab MS-5202, Department of Physics, Columbia University, New York, NY 10027, USA}
\address[4]{Nuclear Science Division, MS 70R0319, Lawrence Berkeley National Laboratory, Berkeley, California 94720 USA}
\address[5]{Institute of Particle Physics, Central China Normal University, Wuhan 430079, China}

\begin{abstract}
First calculated results with the new HIJING++ are presented for identified hadron production in high-energy heavy ion collisions. The recently developed HIJING++ version is based on the latest version of PYTHIA8 and contains all the nuclear effects has been included in the HIJING2.552, which will be improved by a new version of the shadowing parametrization and jet quenching module. Here, we summarize the major changes of the new program code beside the comparison between experimental data for some specific high-energy nucleus-nucleus collisions.
\end{abstract}
\end{frontmatter}

\section{Introduction}
\label{sec:intro}

The original HIJING~\cite{HIJING} (Heavy Ion Jet INteraction Generator) Monte Carlo model was developed by M. Gyulassy and X.-N. Wang with special emphasis on the role of minijets in proton-proton (pp), proton-nucleus (pA) and nucleus-nucleus (AA) reactions at collider energies in a wide range from 5 GeV to 2 TeV. The original program itself is written in FORTRAN, and is based on the FORTRAN version of PYTHIA (version 5)~\cite{PYTHIAv5}, ARIADNE~\cite{ARIADNE}, and the CERNLIB package PDFLIB~\cite{CERNLIB}. This program is today still the most-widely used particle event generator applied for high-energy heavy-ion collisions both in the theoretical model tests and experimental simulations. The main features embedded in the original HIJING are
\begin{itemize}
\item  Soft beam jets are modeled by diquark-quark strings with gluon kinks along the lines of the Lund FRITIOF~\cite{FRITIOF} and dual parton model (DPM). In addition, multiple low-$p_T$ exchanges among the endpoint constituents are included to model initial state interactions.
\item Multiple minijet production with initial and final state radiation is included along the lines of the PYTHIA model in an eikonal formalism.
\item Exact, diffuse nuclear geometry is used to calculate the impact parameter dependence of the number of inelastic processes.
\item An impact-parameter-dependent parton structure function is introduced to study the sensitivity of observables to nuclear shadowing, especially of the gluon structure functions.
\item Gluon radiation from the strings is included according to the ARIADNE program code.
\item Transverse momentum exchange of the created particles simulates the Cronin peak.
\item A simple model for jet quenching is included to enable the study of the dependence of moderate and high-$p_T$ observables on an assumed energy loss of partons traversing the produced dense matter.
\end{itemize}

In 2015 Authors started a project in the international Budapest\,--\,Wuhan\,--\,Berkeley triangle to develop a new version of the HIJING, based on a modern programming language. The aim was to obtain speedup of the program code and update recent physics both on nucleon-nucleon (PYTHIA) and nuclear level. Here we summarize the status of the development and present preliminary results as status report.

\section{Main objectives}
\label{sec:objects}

The HIJING event generator is based on PYTHIA, ARIADNE and PDF libraries, and mainly used for event generation in experimental environment for baseline estimation. Since the todays programming techniques shifted to C++ based programming, the new generation of PYTHIA and PDF libraries are written in this language, furthermore, the experimental platforms are also shifting to C++ (e.g. AliRoot~\cite{ALIROOT}), it is demanding to upgrade the HIJING accordingly. Hence, the main objectives are:
\begin{itemize}
\item to write a genuine, C++ based, modular event generator,
\item to include the most recent public packages (e.g. PYTHIA8~\cite{PYTHIA}, LHAPDF6~\cite{LHAPDF}),
\item support modularity: 
  \begin{itemize}	
	\item possibility of inclusion/change to new theories, alternative processes (e.g. DIPSY~\cite{DIPSY}, Gunion-Bertsch radiation~\cite{Gunion-Bertsch}),
	\item possibility of alternative finite state processes (jet quenching models~\cite{GLV}, etc.),
  \end{itemize}
\item introduce compatibility with experimental analysis platforms (AliRoot),
\item clean and upgrade of the HIJING model itself (e.g. $Q^2$ dependent shadowing~\cite{HOPPET,USHOPPET}),
\item prepare the code for parallel architectures.
\end{itemize}
Since C++ is an object oriented language, it is more suitable for parallelization, being a longer term objective, especially as the evolution of the compilers provide more-and-more efficient computational platform or architecture usage. 

\section{The HIJING++ program}
\label{sec:hijingcode}

Since one of the basic components of HIJNG, the PYTHIA event generator switched to C++ in 2006, it was natural to make a similar change with HIJING. Furthermore, PYTHIA has already developed the appropriate classes to handle elementary collisions, hence the HIJING++ class is positioned in the PYTHIA8 namespace, using its functions to read configuration files (XML), extending parameters with HIJING parameters. However, use of PDF's is tricky, since HIJING is using impact parameter dependent shadowing for nuclei, and the PDF class was extended in order to incorporate masses of nuclei, and the impact parameter of the choosen nucleon pair. The structure may led us even to replace PYTHIA's pseudo-random number generator (PRNG) by a faster GPU-based one~\cite{Barnafoldi-NagyEgri}. 
\begin{figure}[h]
\begin{center}
\includegraphics[width=0.95\linewidth]{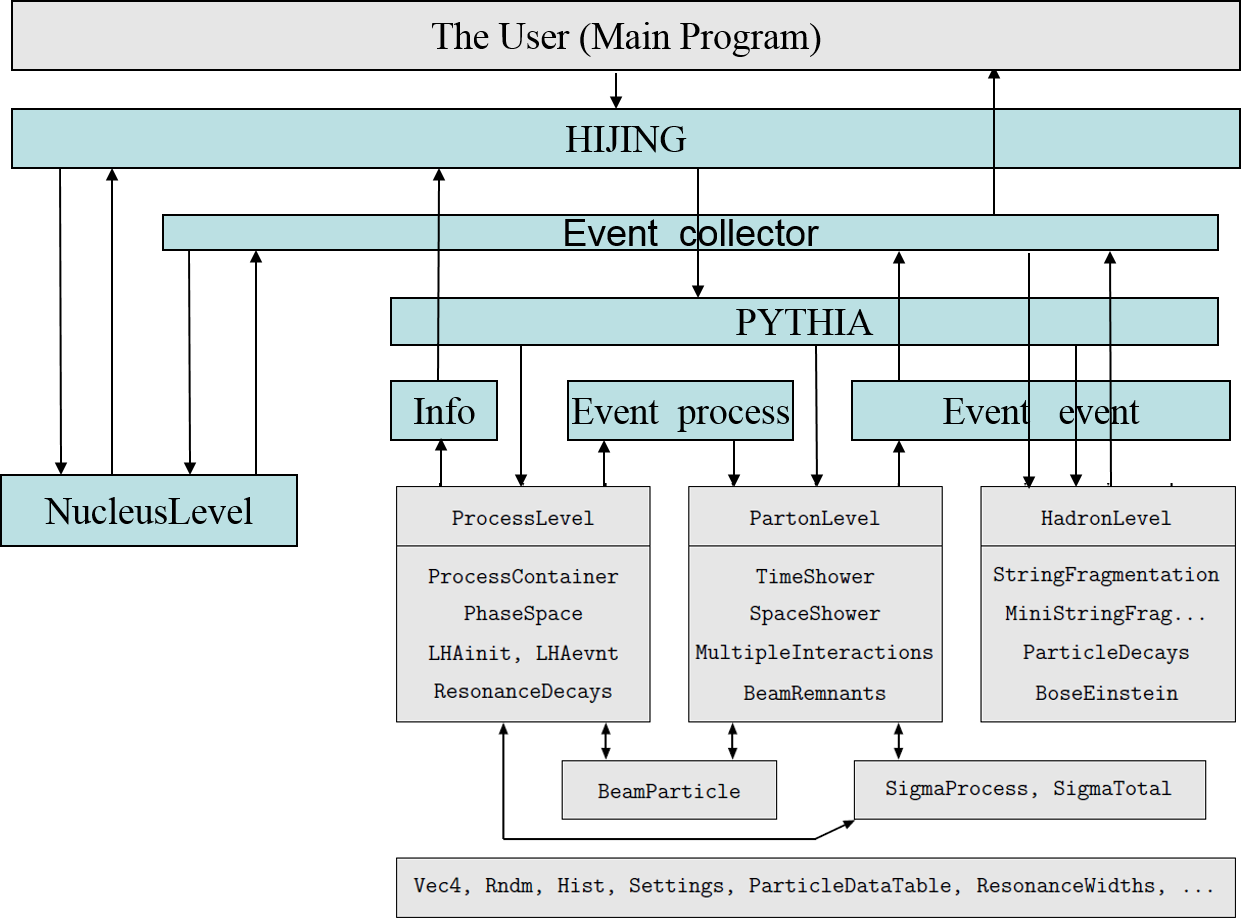}
\label{fig:hijing-struc}
\caption{The structure of the HIJING++ program code.}
\end{center}
\end{figure}

The main structure of the code is presented in Fig.~\ref{fig:hijing-struc}, where colored boxes represent the newly-included {\tt Hijing} class and modifications neglecting cross-links. The {\tt Hijing} class contains all the physics were coded in the FORTRAN subroutines, based on the latest version of HIJING version 2.552~\cite{HIJING2}. Due to the object oriented being of the C++, the original structure was optimized for modularity and future compiler's with improved parallel supports. Technically in {\tt Hijing} processes are ordered in a class hierarchy, common block were changed by class variables and processes are called through object functions. New sub-classes are:
\begin{itemize}
\item {\tt HardCollisions}: for hard $2 \to 2$ processes;
\item {\tt SoftScatter}: for handling soft interactions;
\item {\tt Fragmentation}: based on Lund string model;
\item {\tt NucleonLevel}: for high-energy nuclear effects.
\end{itemize}

\section{Status of the project and the first results}
\label{sec:results}

The first version of C++ code, namely HIJING++ (version 3.1) is ready now for test. The physical models of this recent version is based on PYTHIA8, and HIJING2.552, while the soft physics is still the original ARIADNE version 4 based one. Hence, this version includes all the new physical processes included in the most recent versions of both codes. Here, we present the first physics results of the new HIJING++ version. 

\subsection{HIJING++ results in proton-proton collisions}
\label{sec:hijing-pp}

\begin{figure}[h]
\begin{center}
\includegraphics[width=1.0\linewidth]{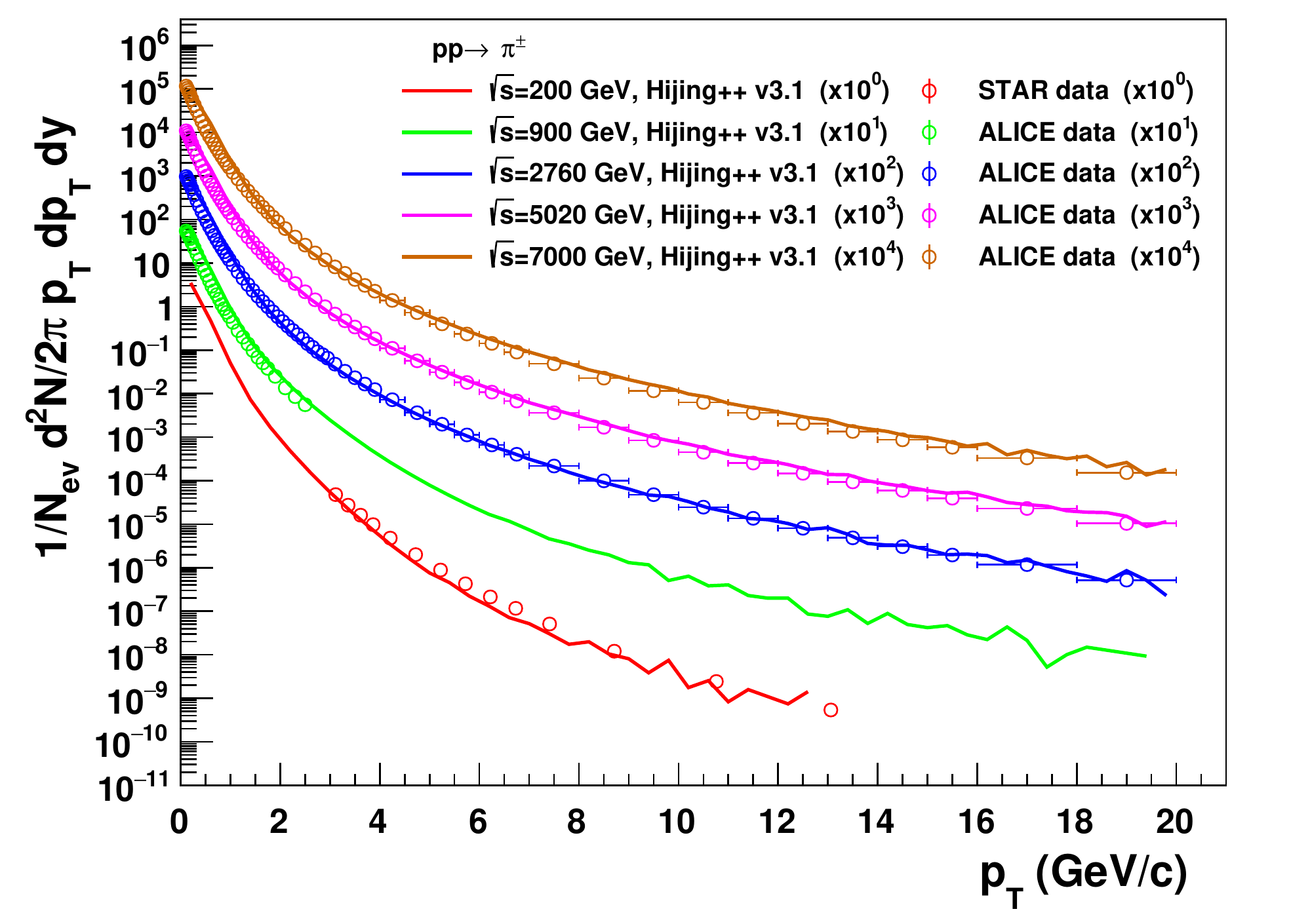}
\label{fig:pp-spectra}
\caption{First preliminary results of HIJING++ compared to the measured charge-averaged pion yield data in proton-proton collisions at 200 GeV~\cite{artic:200gevstar}, 900 GeV~\cite{artic:09tevalice}, 2.76 TeV~\cite{artic:276tevalice}, 5.02 TeV~\cite{artic:5tevalice}, and 7 TeV~\cite{artic:7tevalice,artic:7tevalice2} c.m. energies. Spectra were shifted for better visibility.}
\end{center}
\end{figure}

In Figure~\ref{fig:pp-spectra} we plotted the charge-averaged identified pion yields, calculated and compared to experimental data in proton-proton collisions at 200 GeV~\cite{artic:200gevstar}, 900 GeV~\cite{artic:09tevalice}, 2.76 TeV~\cite{artic:276tevalice}, 5.02 TeV~\cite{artic:5tevalice}, and 7 TeV~\cite{artic:7tevalice,artic:7tevalice2} center-of-mass energies. We used 10 M events for all energy values in the $|\eta|<0.8$ range. Note, kaon and proton spectra present the same agreement. 

First comparison between measured spectra {\sl dots} and the HIJING++ calculated {\sl curves} shows nice overlap at high accuracy, especially at the highest center-of-mass energies. At lower energies and small momenta, where soft processes play the role, the data over theory start to deviate, which suggest to apply the proper PYTHIA tune in the tests for further development phases. 

\subsection{HIJING++ results in proton-nucleus collisions}
\label{sec:hijing-pA}

We tested the HIJING++ code in minimum bias proton-lead (pPb) collision at LHC energy, $\sqrt{s_{NN}}=5.02$ TeV. For the case of the pPb collision we used 5M events and calculated the charge-averaged yield of pion, kaon, and proton productions. Here $-0.5 < \eta <0$ was applied for all hadron species. Nuclear effect are based on the settings adopted from the original HIJING2.552, including impact-parameter dependent nuclear shadowing, multiple scattering, and energy-dependent minijet cutoff as given in Ref.~\cite{HIJING2}.  
\begin{figure}[h]
\begin{center}
\includegraphics[width=1.0\linewidth]{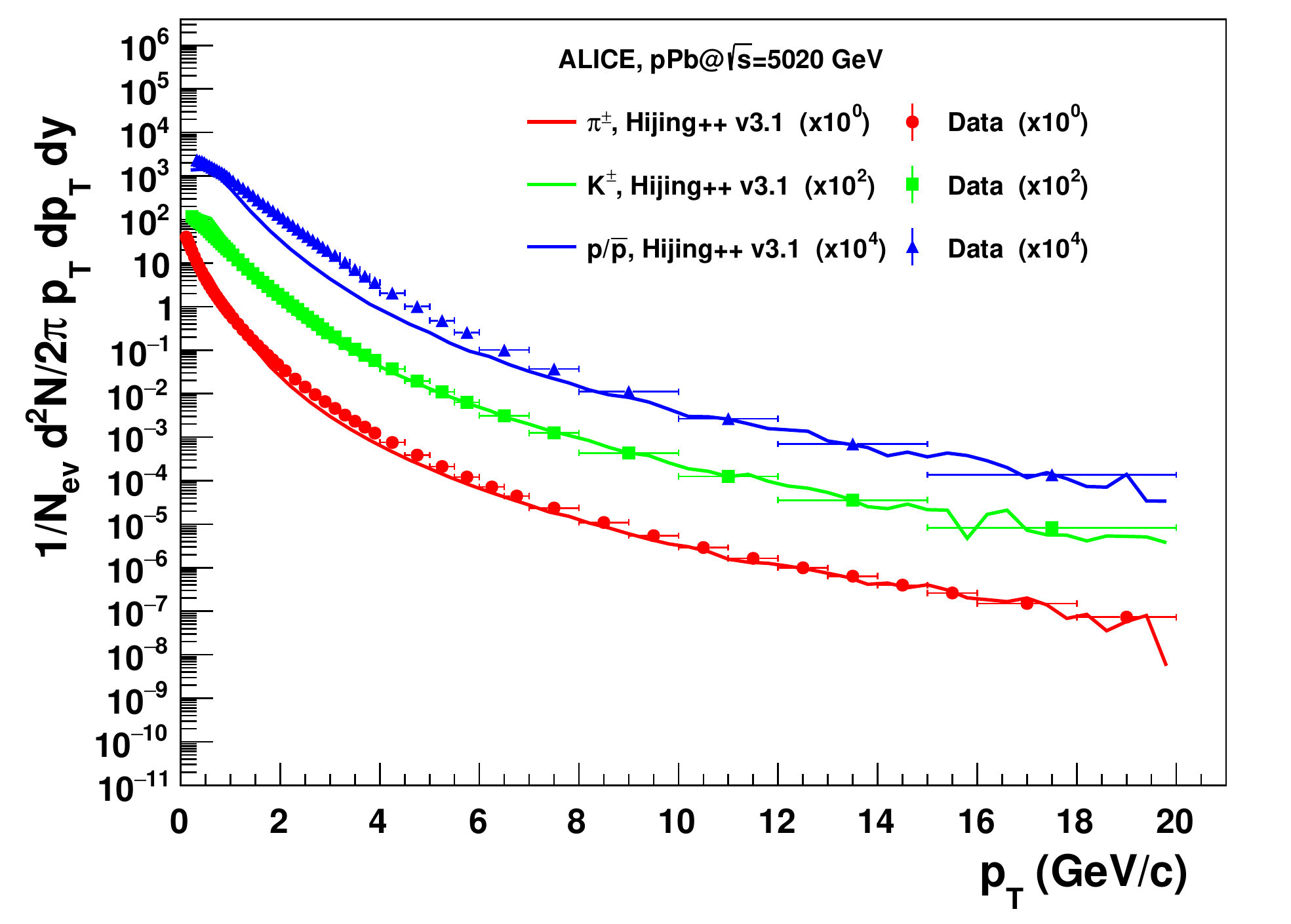}
\label{fig:pA-spectra}
\caption{First preliminary results of HIJING++ on charged-averaged pion, kaon, and proton spectra in proton-lead collisions at $\sqrt{s_{NN}}=5.02$ TeV energy in comparison with ALICE~\cite{artic:5tevalice} data. Note, hadron spectra species were shifted for better visibility.}
\end{center}
\end{figure}

In Figure~\ref{fig:pA-spectra} the results obtained by HIJING++ on charged-averaged pion, kaon, and proton spectra are presented in pPb collisions at $\sqrt{s_{NN}}=5.02$ TeV energy in comparison with ALICE~\cite{artic:5tevalice} data. Kaon and proton spectra were shifted by factors of $10^2$ and $10^4$, respectively for the better visibility.

The calculated spectra show agreement with the identified hadron yield measurement, especially at the 
highest and lowest transverse momentum regimes. In the intermediate region, 2 GeV/$c <p_T< 7$ GeV/$c$ --- where high-energy nuclear effects take place --- the theoretical curves deviate from the experimental data. In the present status of the HIJING++ development we are now on, to identify and fine-tune the strength of these effects, especially according to the the latest, available, high-accuracy RHIC and LHC data. This is a challenging task, since in these phenomenological models we need to satisfy all previous and recently measured data in several manners. On the other hand we need to synchronize the included sub-nuclear theoretical features with the nuclear phenomenology to avoid double counting of effects.

\section{Ongoing developments and future perspectives}
\label{sec:devels}

In parallel to the test phase of the coding, a sophisticated theoretical development of the model is ongoing. Here, we list some of these updates and features, which will present in the final, public code: 

\begin{itemize}
\item The original HIJING shadowing function will be replaced by a scale-dependent nuclear shadowing. The $Q^2$-evolution is based on the original HIJING shadowing and the HOPPET~\cite{HOPPET} evolution code. This modification softens the shadowing effect for larger $Q^2$ values according to the lack of nuclear modification measurements~\cite{USHOPPET}.

\item A contemporary jet energy loss module is under development. This module is enable users to include various jet-quenching models in the future, such as we include Gyulassy\,--\,L\'evai\,--\,Vitev (GLV)~\cite{GLV}.  

\item The run-time of the HIJING++ is comparable to the previous, FORTRAN based version, however, this new version is suitable for parallel architectures, which may speed up the calculations considerably. The present code structure is capable to provide thread-level and/or MPI parallelization. 

\item Since the new version of the code is written in the modular C++, it is natural to include it to huge detector simulation frameworks, like ALICE's AliRoot~\cite{ALIROOT}. Due to the modularity, merging of these simulation frameworks can reduce their volume and the number of cross links, thus simulations become more memory consumable.  

\item Since the code is written completely in C++ a (partial) parallel-platforms supported version of the program is planned to develop. The preliminary tests showed, that changing the random number generator to a GPU-based version, can result in a slight increase of the speed. Optimalization for various multi-core and parallel architectures is on the wish list. 
\end{itemize}

\section{Summary and outlook}
\label{sec:summary}

Authors summarize here the first results calculated by the developed Monte Carlo heavy ion jet interaction generator, HIJING++. The well-know and widely-used FORTRAN-based HIJING, were rewritten
in a modern and evolving programming language C++. Adopting the structure of the PYTHIA8, the {\tt Hijing} class were invented, including the nuclear mechanism modeled in the original HIJING version. We presented the comparison of charge-averaged identified hadron yields in proton-proton collisions from 200 GeV to 7 TeV c.m. energies and at LHC energy, $\sqrt{s_{NN}}=5.02$ TeV, in minimum bias proton-lead collisions in comparison to experimental data.   

Present milestone aims to communicate the status of this software development, moreover, give perspectives for the forthcoming applicabilities and features of the soon-to-be-publised open source HIJING++ for the next generation of heavy-ion collision measurement, simulations, and facilities at future colliders.

\section*{Acknowledgements}

This work was supported by the Hungarian-Chinese cooperation grant No T\'eT 12 CN-1-2012-0016 and No.
MOST 2014DFG02050, Hungarian National Research Fund (OTKA) grants NK106119 and K120660. Author G.G. Barnaf\"oldi also thanks the J\'anos Bolyai Research Scholarship of the Hungarian Academy of Sciences. We acknowledge the support of the Wigner GPU laboratory.

\nocite{*}
\bibliographystyle{elsarticle-num}
\bibliography{jos}

\begin{thebibliography}{00}
\bibitem{HIJING} X.N. Wang, M. Gyulassy,  Phys. Rev. {\bf D44}, 3501  (1991).

\bibitem{HIJING2} W.T. Deng, X.N. Wang, R. Xu, Phys. Rev. {\bf C83}, 014915 (2011).

\bibitem{PYTHIA} T. Sj\"ostrand, Comput. Phys. Commun. {\bf 191}, 159 (2015).

\bibitem{ALIROOT} AliRoot: http://aliweb.cern.ch/Offline/AliRoot/Manual.html (2017)

\bibitem{PYTHIAv5} T.~Sjostrand,
  Comput.\ Phys.\ Commun.\  {\bf 82}, 74 (1994).

\bibitem{ARIADNE} L. L\"onnblad, Comput. Phys. Comm. {\bf 71}, 15 (1992).

\bibitem{CERNLIB} CERNLib:  https://cernlib.web.cern.ch/cernlib/ (2017)

\bibitem{FRITIOF}  B.~Nilsson-Almqvist and E.~Stenlund,
  Comput.\ Phys.\ Commun.\  {\bf 43}, 387 (1987).

\bibitem{LHAPDF} LHAPDF6: http://lhapdf.hepforge.org/  (2017)

\bibitem{HIJINGsh} X.N. Wang, Phys. Rev.{\bf C61}, 064910 (2001).

\bibitem{DIPSY} C. Flensburg,  Prog.Theor.Phys.Suppl. {\bf 193}, 172 (2012). 

\bibitem{Gunion-Bertsch} J.F. Gunion and G. Bertsch, Phys. Rev.{\bf D25}, 746 (1982)

\bibitem{GLV} M.Gyulassy, P.Levai, I.Vitev, Phys .Rev. Lett. {\bf 85}, 5535 (2000).

\bibitem{HOPPET} A. Vogt, S. Moch, J.A.M. Vermaseren, Nucl. Phys. {\bf B691}, 129 (2004) 

\bibitem{USHOPPET}G. Ma, G.G. Barnaf\"oldi, Weitian Deng, Sz. Harangoz\'o, G. Papp, X-N. Wang, B-W. Zhang, DGLAP-evolved Shadowing Parametrization for Simulating High-energy Nucleus-Nucleus Collisions in HIJING, ({\sl in preparation})

\bibitem{Barnafoldi-NagyEgri} G.G. Barnaf\"oldi, M.F. Nagy-Egri, GPU-based PRNG for Monte Carlo Particle Event Genarators ({\sl in preparation})

\bibitem{artic:200gevstar}
STAR Collaboration. {\em Phys. Rev. Lett.} {\bf 2012}, {\em 108}, 072302.

\bibitem{artic:09tevalice}
ALICE Collaboration. {\em Eur. Phys. J C}  {\bf 2010}, {\em 71}, 1655.

\bibitem{artic:276tevalice}
ALICE Collaboration. {\em Phys. Lett. B}  {\bf 2014}, {\em 736}, 196-207.

\bibitem{artic:5tevalice}
ALICE Collaboration. {\em Phys. Rev. C} {\bf 2015}, {\em 91}, 064905.

\bibitem{artic:7tevalice}
ALICE Collaboration. {\em Eur. Phys. J. C}  {\bf 2015}, {\em 75(5)}, 226.

\bibitem{artic:7tevalice2}
ALICE Collaboration. {\em Phys. Lett. B}  {\bf 2016}, {\em 760}, 720.


\end{thebibliography}


\end{document}